\documentclass[]{spie}  
\pdfoutput=1

\usepackage{amsmath,amsfonts,amssymb}
\usepackage{graphicx}
\graphicspath{{images/}} 
\usepackage[colorlinks=true, allcolors=blue]{hyperref}
\usepackage{xcolor}
\usepackage{subcaption}

\begin{document} 
\title{Quantifying the contamination from nebular emission in NIRSpec spectra of massive star forming regions.}

\author[a]{Ciaran R. Rogers}
\author[b]{Guido De Marchi}
\author[b]{Giovanna Giardino}
\author[a]{Bernhard R. Brandl}
\author[c]{Pierre Feruit}
\author[d]{Bruno Rodriguez}
\affil[a]{Leiden Observatory, PO Box 9513, 2300 RA, Leiden, The Netherlands}
\affil[b]{European Space Agency, Keplerlaan 1, PO Box 299, 2200 AG, Noordwijk, The Netherlands}
\affil[c]{European Space Astronomy Centre, Camino bajo del Castillo, s/n Urbanización Villafranca del Castillo Villanueva de la Cañada E-28692 Madrid, Spain}
\affil[d]{Centre for Astrobiology, Unidad María de Maeztu, Instituto Nacional de Técnica Aeroespacial, Ctra de Torrejón a Ajalvir, km 4, 28850 Torrejón de Ardoz, Madrid, Spain}

\authorinfo{Further author information: (Send correspondence to C.R.R.)\\C.R.R.: E-mail: rogers@strw.leidenuniv.nl\\ G.M.: E-mail: gdemarchi@cosmos.esa.int, G.G.: E-mail: ggiardin@cosmos.esa.int, B.R.B.: E-mail: brandl@strw.leidenuniv.nl, P.F.: E-mail: pierre.ferruit@esa.int, B.R.: E-mail: brodriguez@cab.inta-csic.es}

\pagestyle{empty} 

\maketitle

\begin{abstract}
The Near InfraRed Spectrograph (NIRSpec) on the James Webb Space Telescope (JWST) includes a novel micro shutter array (MSA) to perform multi object spectroscopy. While the MSA is mainly targeting galaxies across a larger field, it can also be used for studying star formation in crowded fields. Crowded star formation regions typically feature strong nebular emission, both in emission lines and continuum. In this work, nebular emission is referred to as nebular contamination. Nebular contamination can obscure the light from the stars, making it more challenging to obtain high quality spectra. The amount of the nebular contamination mainly depends on the brightness distribution of the observed `scene'. Here we focus on 30 Doradus in the Large Magellanic Cloud, which is part of the NIRSpec GTO program. Using spectrophotometry of 30 Doradus from the Hubble Space Telescope (HST) and the Very Large Telescope (VLT)/SINFONI, we have created a 3D model of the nebular emission of 30 Doradus. Feeding the NIRSpec Instrument Performance Simulator (IPS) with this model allows us to quantify the impact of nebular emission on target stellar spectra as a function of various parameters, such as configuration of the MSA, angle on the sky, filter band, etc. The results from these simulations show that the subtraction of nebular contamination from the emission lines of pre-main sequence stars produces a typical error of $0.8\%$, with a $1\sigma$ spread of $13\%$. The results from our simulations will eventually be compared to data obtained in space, and will be important to optimize future NIRSpec observations of massive star forming regions. The results will also be useful to apply the best calibration strategy and to quantify calibration uncertainties due to nebular contamination.  
\end{abstract}

\keywords{Astronomical instrumentation, James Webb Space Telescope, Instrumentation: Spectrographs, Nebular contamination, Micro Shutter Assembly, SPIE Proceedings}

\section{INTRODUCTION}
\label{sec:intro}  

The James Webb Space Telescope is equipped with four science instruments, one of which is the Near InfraRed Spectrograph (NIRSpec). Operating between $0.6 - 5 \mu m$, NIRSpec has been designed to probe the high red-shift universe, to try to understand the formation of galaxies during the era of reionisation. The versatility of NIRSpec makes it possible for the instrument to also be used to study the formation and evolution of stars and planets, as well as the detection of water and other molecules in exoplanet atmospheres.\\
To probe these questions, NIRSpec is capable of performing three types of spectroscopy, namely: long slit spectroscopy, integral field spectroscopy, and multi-object spectroscopy (MOS). MOS mode utilises the novel Micro Shutter Array (MSA), designed and produced by the Goddard Space Flight Centre. The MSA consists of roughly 250000 apertures, each about $100 \mu m$ in width, and $200 \mu m$ in length. By means of an electromagnetic arm, the apertures can be configured open, or closed, depending on the requirements of the observation. The MOS mode is well suited for studying the formation and evolution of galaxies, with its wide $3'.6 \times 3'.4$ field of view. It is also a great tool for studying star formation in crowded fields, thanks to its highly configurable array of micro shutters.\\ 
When studying crowded fields, unwanted light is often detected from the environment around the targets. This unwanted emission, here called nebular contamination, becomes interwoven with light from the targets, resulting in superimposed emission lines in the target spectra, as well as an increase in the overall continuum level. Nebular emission comes in two varieties: continuum emission and line emission. The radiative processes behind continuum emission in H II regions are free-free emission (bremsstrahlung), bound-free emission, and two-photon emission. Line emission is caused by electronic transitions occurring between bound states in atoms as well as fine-structure line transitions \cite{byler2017nebular}. These emission lines become blended with the stellar absorption features, resulting in what is known as `filling in' \cite{walborn1990contemporary}. As a result, studying the absorption lines of stellar spectra with nebular contamination is complicated, as the true equivalent width and radial velocities cannot be directly determined. The same is true for studying pre-main sequence (PMS) stellar emission lines. In this case, the observed emission line is a combination of the stellar component and the nebular component. Disentangling one from the other can require careful background subtraction, which is the focus of this paper. The presence of strong nebular emission may also reduce the contrast of observations. Dim sources may be outshone by the nebular background, rendering them barely detectable or completely invisible. There is also contamination from the zodiacal light, which originates from dust in the solar system. Typically this is on the order to $10-100$ times less bright than HII regions, and so its impact is minimal here.\\
In this paper, we try to determine the impact that nebular emission will have on NIRSpec observations in the starburst region 30 Doradus in the Large Magellanic Cloud. 30 Doradus will be observed as part of the NIRSpec GTO program. One of the goals of this program is to understand the role that metallicity plays in star formation. The equivalent width (EW) of emission lines from PMS stars can be used to measure their mass accretion rate \cite{dahm2008spectroscopic}. Nebular contamination changes the EW of these lines, and hence the successful removal of the nebular component is crucial in accurately determining the accretion rate. In order to gauge the effect that nebular contamination will have on observations of PMS stars in 30 Doradus, a model nebular emission spectrum was created, and run through the Instrument Performance Simulator (IPS) along with dozens of stellar models in order to simulate a real NIRSpec observation with contamination from the background. The dispersion and grating selection used by the IPS was G235M/F170LP, which refers to the medium resolution setting (R=1000) of NIRSpec, covering a wavelength range from $1.66-3.07 {\mu}m$. In section 2 a description of the model background spectrum is given. Section 3 describes the simulation of a NIRSpec observation. In section 4, the results from the IPS simulation are presented and discussed.    

\section{Creating the model nebular emission spectrum}

\subsection{Measuring the nebular emission in 30 Doradus}
\begin{figure}[h]
    \centering
    \includegraphics[width=0.7\linewidth]{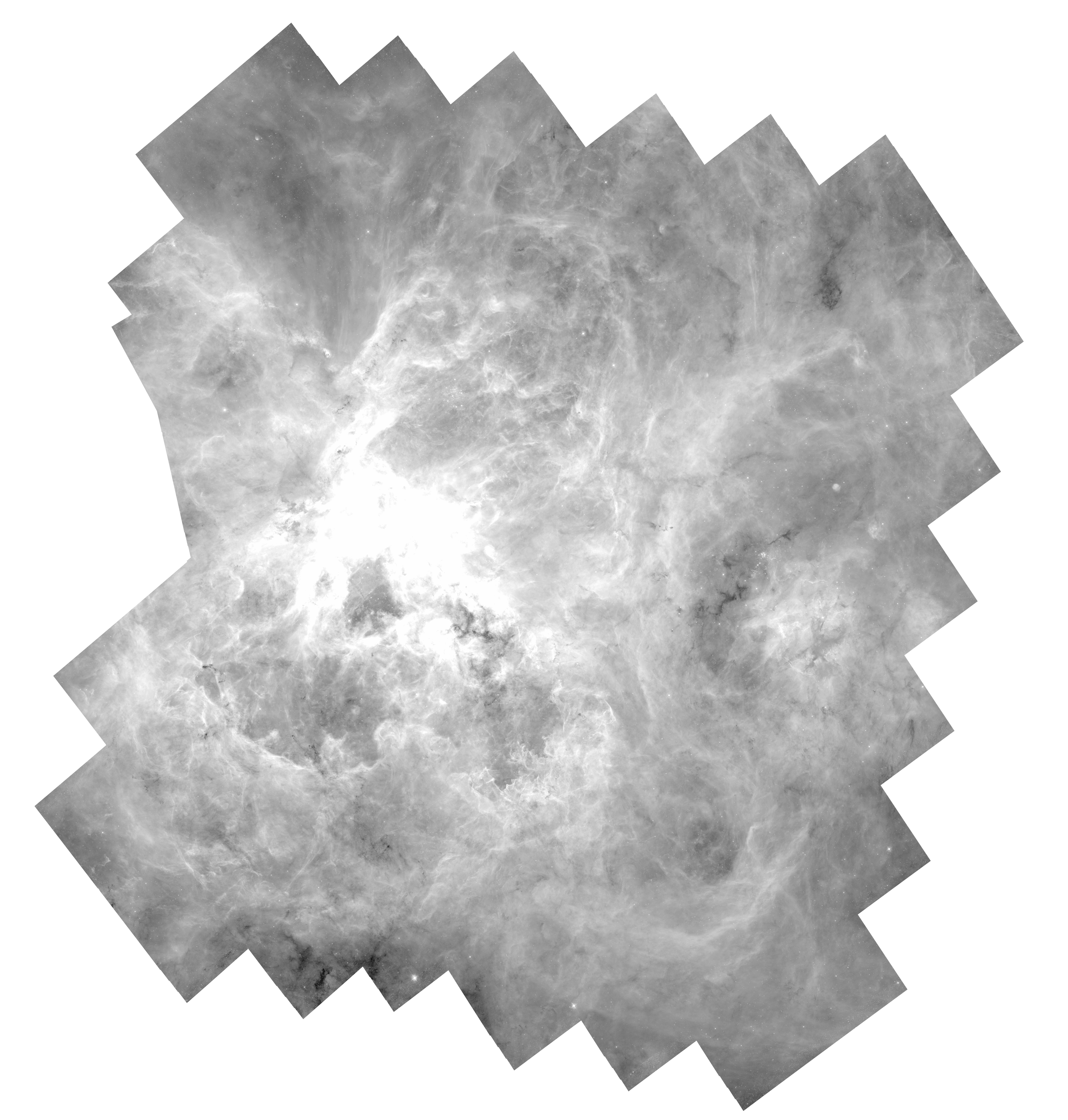}
    \caption{Image of 30 Doradus viewed through the $F658N$ filter using the ACS on board HST. The region in the centre left of the image that resembles a Christmas tree was used to determine the instrumental background \cite{sabbi2013hubble}.}
    \label{fig:30_dor_halpha}
\end{figure}

\begin{figure}[h]
    \centering
    \includegraphics[width=0.7\linewidth]{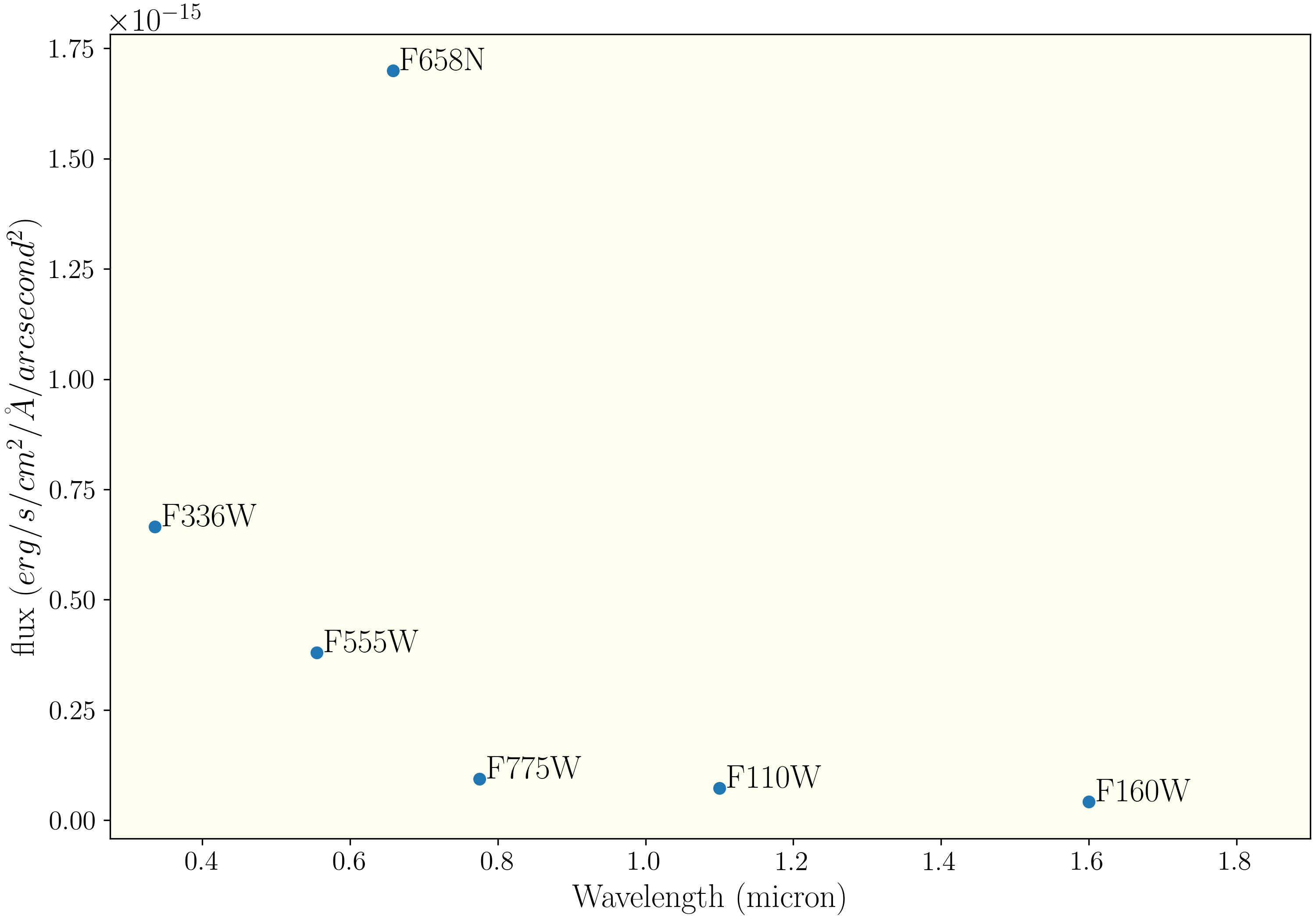}
    \caption{Nebular emission brightness versus wavelength in the core of 30 Doradus.}
    \label{fig:photometric_plot}
\end{figure}

In order to determine the impact that nebular emission will have on observations of dense star forming regions with NIRSpec, a model spectrum was created, using drizzled mosaic photometry collected by HST's Wide Field Camera 3 (WFC3) and Advanced Camera for Surveys (ACS) in 2013. These data are described in detail in \cite{sabbi2013hubble} and \cite{sabbi2016hubble}. Of these observations of the central region of 30 Doradus, six photometric filters were used here. WFC3 observed the region through filters: $F336W, F110W, F160W$ and ACS used filters: $F555W, F658N, F775W$. WFC3 has a field of view of $160" \times 160"$ and ACS has a field of view of $202" \times 202"$. An image of 30 Doradus viewed through the narrow $H_\alpha$ filter at $658 nm$ is shown in Figure \ref{fig:30_dor_halpha}.\\ 
The approach to determine the nebular background level from the data was as follows. For each photometric image, a virtual aperture was created, and placed in the core of 30 Doradus, away from any stars in order to ensure that only nebular light was being measured. This region was chosen because of its high surface brightness. This allowed us to simulate a `worst case scenario' in terms of nebular contamination, providing us with an upper limit to the adverse affect of the nebular emission. The area of the aperture in pixels was chosen to be equal to that of 3 micro shutters (also known as a `mini-slit'). Each micro shutter has an open area of $ 0.20" \times 0.46"$ projected on the sky. The NIRspec detector pixels, which measure $18 /mu m \times 18 /mu m$, have a projected area of $0.103" \times 0.105"$ \cite{ferruit2022near}. This means that a single micro shutter illuminates about 8 pixels, and hence the mini-slit illuminates about 24 pixels. As such, a rectangular aperture was created with dimensions $x = 3$ $pixels$ and $y = 8$ $pixels$, giving a total area of 24 pixels. The aperture was rotated to the roll angle that JWST will have when it observes 30 Doradus as part of the NIRSpec GTO program, namely $336.15168149^{\circ}$. The region of the nebula that fell within this aperture was measured and the median pixel value was determined. This value represented the average brightness of the nebula in that region.\\ 
The area in figure \ref{fig:30_dor_halpha} that resembles the outline of a Christmas tree was used to measure the instrumental background. This region is likely an outflow of material, driven outwards by previous generations of star formation \cite{guido_30dor}. It features lower column densities than the surrounding material and is optically thin. Hence very little of the light detected here comes from the nebula or stars, but predominantly from the instruments themselves, as well as the foreground zodiacal light. The median value of this Christmas tree area was subtracted from the value found in the core region, in order to isolate just the light originating within 30 Doradus. This same process was then repeated for the other filters.\\
The brightness values were initially in electrons s$^{-1}$ and so had to be converted to physical units (in this case: erg cm$^{-2}$ s$^{-1}$ $\AA^{-1}$ arcsec$^{-2}$). To do this, the so called `inverse sensitivity' of the image must be known, which is provided in the FITS header of each image. Inverse sensitivity represents the flux required to produce one count per second on the detector. Using the inverse sensitivity as well as the plate scale, the nebular emission level was converted to physical units. Typically the plate scale of a photometric image is specific to the camera. In the case of the 30 Doradus drizzled mosaic images, the same plate scale was used for all images. This was done so that images of different wavelengths could be compared directly with one another. The plate scale for the data used here was $0.03962"/pixel$. The photometric fluxes in physical units are plotted against wavelength, shown in figure \ref{fig:photometric_plot}.

\subsection{Fitting a spectrum to the data}
\begin{figure}[h]
    \centering
    \includegraphics[width=0.7\linewidth]{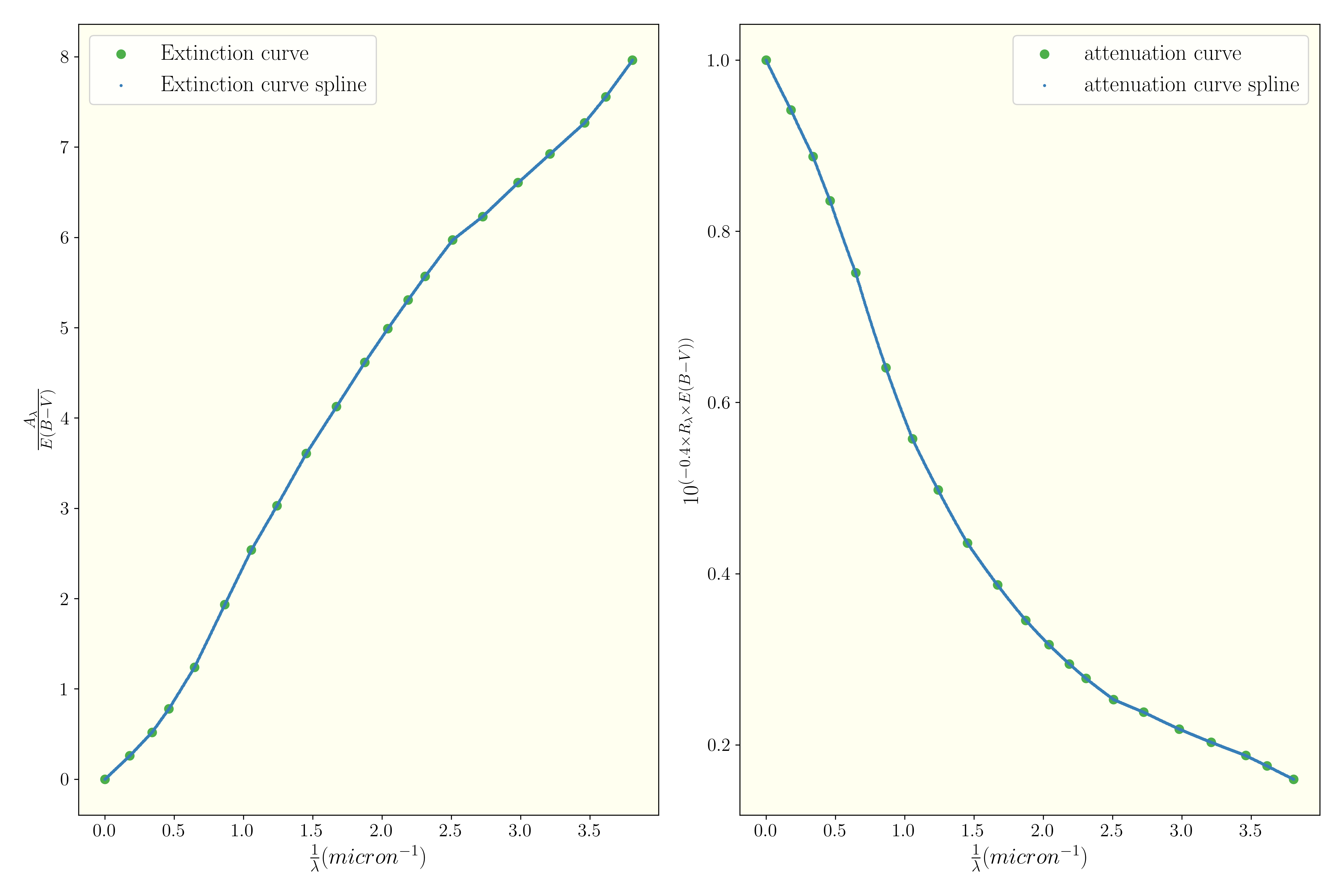}
    \caption{Left: The extinction curve of 30 Doradus. Right: The attenuation curve resulting from the conversion of magnitudes to flux. The Y axis values represent the attenuating factor that is to be multiplied by the flux of the model spectrum for a given wavelength, in order to account for dust extinction.}
    \label{fig:extinction_attenuation_curves}
\end{figure}

\begin{figure}[h]
    \centering
    \includegraphics[width=0.7\linewidth]{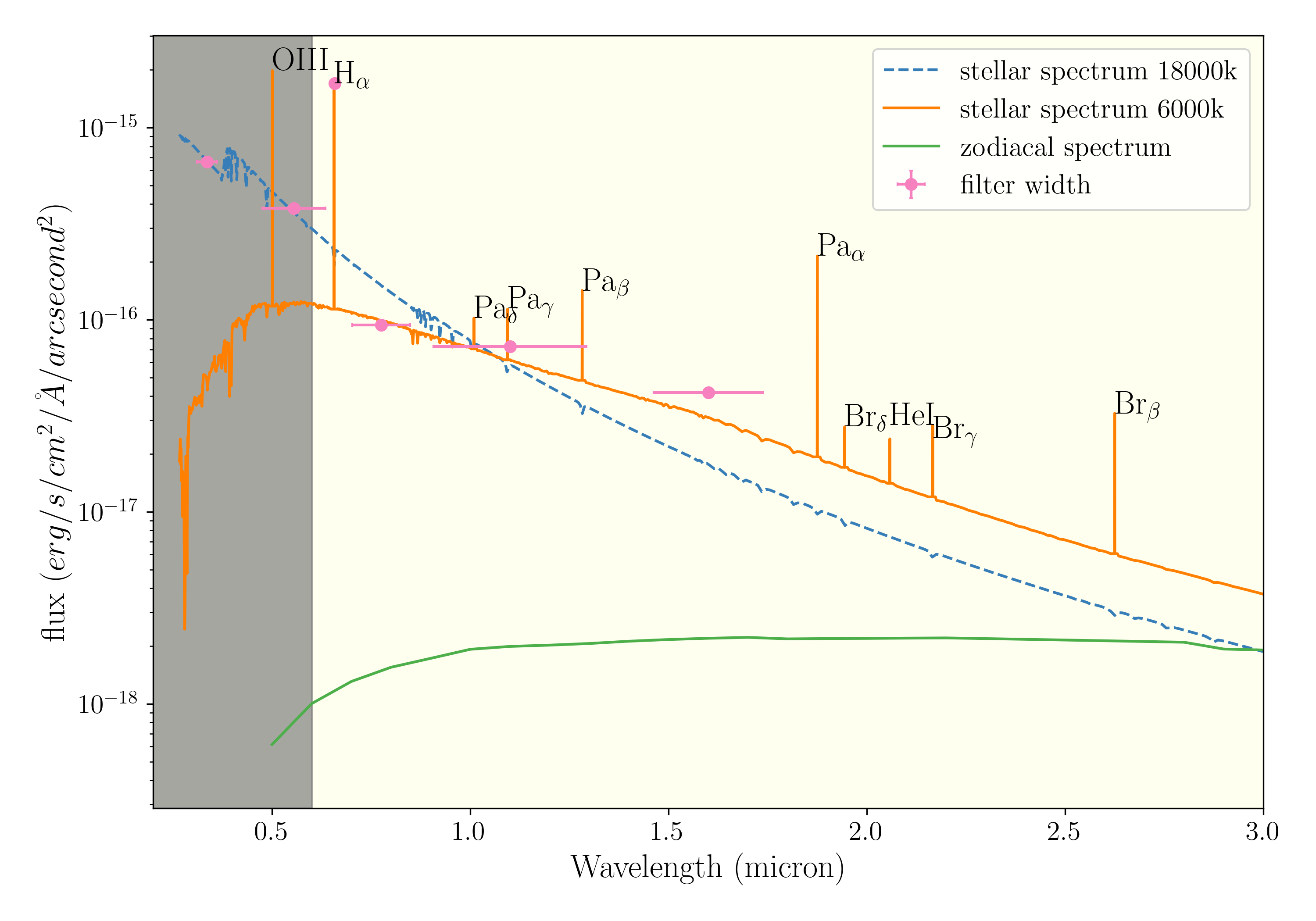}
    \caption{The stellar spectra used to fit the photometric fluxes obtained from HST of 30 Doradus. The pink points are the photometric fluxes. The blue spectrum is that of an 18000K star. The orange spectrum is that of a 6000K star. The green line represents the zodiacal light spectrum, shown here to illustrate the difference in brightness between it and the nebular emission. The shaded region indicates the wavelength range not covered by NIRSpec.}
    \label{fig:reddened_background_spectrum}
\end{figure}

The next step in creating the model background spectrum involved fitting a stellar spectrum to the photometric points. The spectrum of a HII region is the response of the nebula to the ionising photons produced by its OB stars; hence the nebular spectrum will share characteristics with that of the stars that ionise it. However, radiative processes occur in nebulae that can not occur in stars. This results in key differences between stellar and nebular spectra. Despite these differences, within the wavelength range of NIRSpec, stellar spectra are a good approximation. 30 Doradus is an extremely active and bright HII region, powered by many early O and and B stars \cite{melnick198530}. To begin fitting spectra to the photometric points, model stellar spectra from the ATLAS9 Kurucz ODFNEW/NOVER catalogue were used, ranging in effective temperatures $T_{eff}$ from $5000-20000K$ (with a fixed $\log{g} = 4.5 cm/s^2$ and metallicity = $-0.5$) as appropriate for the Large Magellanic Cloud cite{geha1998stellar}. Before fitting could be done, the Hydrogen, Helium and metal emission lines present in 30 Doradus needed to be added to the spectra. The relevant emission lines spanned from about $0.5 - 3 \mu m$ in wavelength based on the dispersion and filter selection.\\
Using the ACS H${_\alpha}$ filter $F658N$, the intensity of the H${_\alpha}$ emission line was derived from the $F658N$ photometric flux. The line itself was modelled as a simple Gaussian with a full width half maximum of $6 \AA$. At this width, none of the lines are resolved by NIRSpec, which is appropriate for nebular emission lines. Based on Osterbrock's case B recombination calculations \cite{osterbrock1981seyfert}, the emission strength of the other Hydrogen series emission lines is simply a fixed ratio of the H${_\alpha}$ emission line strength. Based on this, the intensities of the Paschen and Brackett series emission lines were determined, and these lines added to the spectra. Because the model spectra being used here were main sequence stars, stellar absorption lines were present where the nebular emission lines were being added. These lines were removed by selecting a region around each absorption line, and setting the flux equal to the nearby continuum. Other absorption lines present in the spectra were ignored, as their presence made no difference to the simulations. The strength of the helium emission line was determined by comparing the strength of He I $\lambda 2058.0$ to the nearby Br${_\gamma} \lambda 2166.0$. For this, IFU observations of 30 Doradus obtained with the SINFONI instrument were used. These datacubes, obtained from the ESO archives, contained both the He I $\lambda 2058.0$ and Br${_\gamma} \lambda 2166.0$ lines. The ratio of these two lines was found to be equal to $0.61$. This ratio was used to set the intensity of the helium line in the model spectrum. Finally the prominent O III $\lambda 5006.8$ emission line was added. The intensity of this line was taken directly from optical observations of 30 Doradus in \cite{crowther2018dissecting}. Though this line falls outside the wavelength range of NIRSpec, it was needed in order to properly fit the spectrum to the photometric points. In the presence of strong emission lines (several 10s of times brighter than the continuum), photometric points will move noticeably away from the continuum. This is simply because the flux measured in a filter is due not only to continuum, but also due to emission lines.\\ 
The model spectra also needed to be reddened, due to extinction from the dust in 30 Doradus. The spectra were attenuated using the extinction curve of 30 Doradus determined by \cite{guido_30dor}, using the relationship: $10^{(-0.4*A_\lambda)}$, where $A_{\lambda}$ is the extinction in magnitudes at various wavelengths. The extinction curve and attenuation curve are shown in figure \ref{fig:extinction_attenuation_curves}. The attenuated spectra were then multiplied by a scaling factor in order to bring them to the same brightness as the photometry. By incrementally raising the surface temperature, the best fitting model could be found. It became apparent during this step that no single stellar spectrum could provide a reasonable fit to all of the measured photometric points. The blue/visible points were best fit by a mid-B type spectrum, while the red and IR points resembled a much cooler spectrum. Ultimately a spectrum with $T_{eff} = 6000K$ was used to fit the photometric points at wavelengths longer than $F658N$, while a spectrum with $T_{eff} = 18000K$ was used to fit the photometric points at shorter wavelengths. It is expected that multiple stellar models are needed to fit the spectrum of a HII region, because of the different radiative processes involved, including the 2-photon process and dust emission, which are not considered in these stellar models. Figure \ref{fig:reddened_background_spectrum} shows the best fitting spectra to the photometric data. The grey shaded region highlights the wavelengths that fall outside of NIRSpec's wavelength range. These data were still important in understanding the radiative processes occurring in 30 Doradus, and helped inform the choice of spectra.

\section{Simulating a Multi-Object Spectroscopy observation}
\subsection{Configuring the MSA}
\begin{figure}[h]
    \centering
    \includegraphics[width=0.7\linewidth]{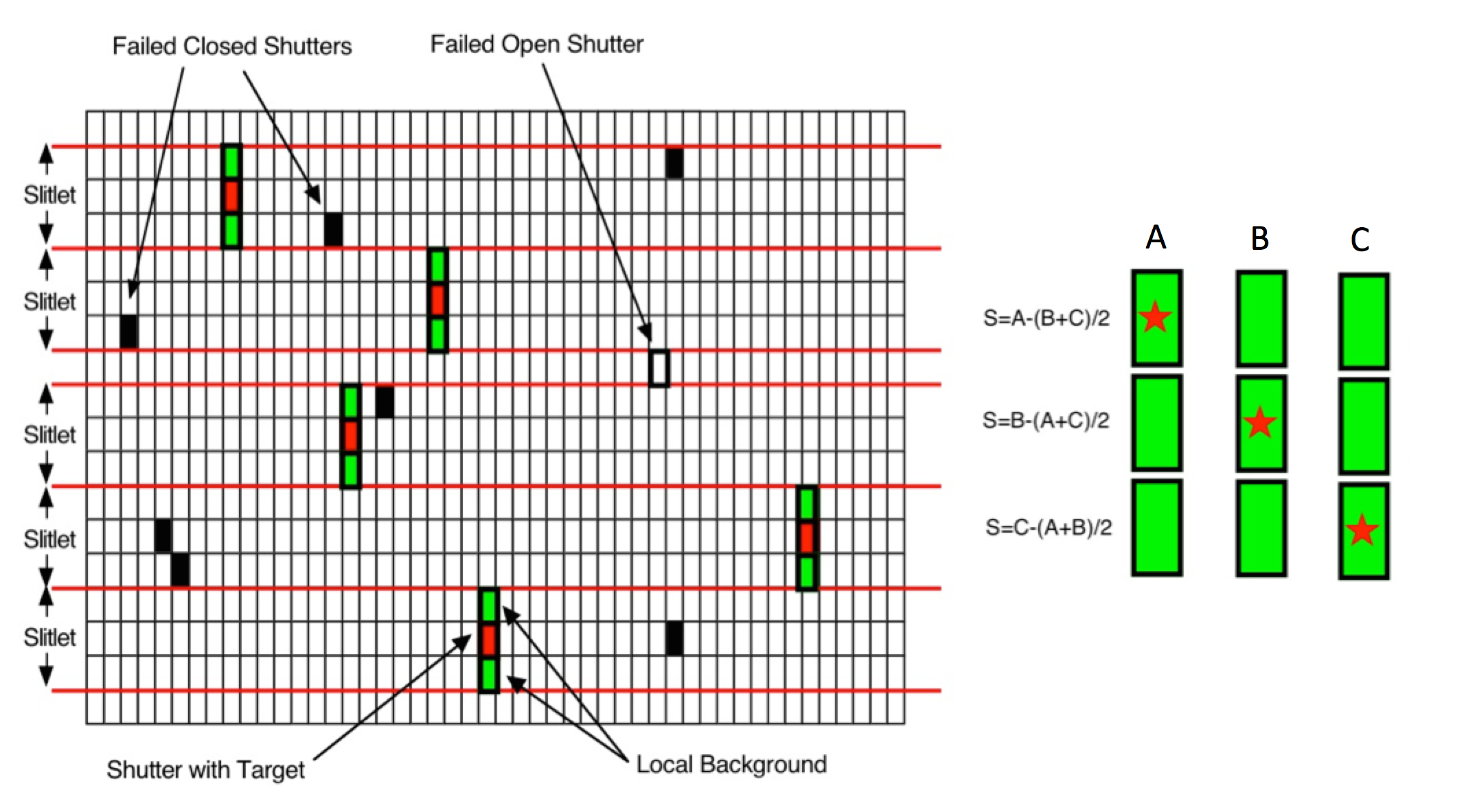}
    \caption{The nodding pattern used in this simulation is illustrated above, showing a section of the MSA, highlighting how a target moves from one shutter to another during the 3 nodding movements.}
    \label{fig:MSA_sky_strategy}
\end{figure}

A MOS observation with NIRSpec requires careful planning in order to maximise the number of targets that can be observed \cite{ferruit2022near}. The planning of the MSA configuration is typically done via the Microshutter Planning Tool (MPT) \cite{ferruit2022near}. The planning requires knowledge of the roll angle of JWST, so that the angle of the micro shutters with respect to the field is known. The MSA is then configured, selecting which shutters shall be open, and which shutters shall remain closed. This depends on the location of the targets. A MOS observation can measure as many as 100 stellar spectra simultaneously \cite{ferruit2022near}. When observing a star in MOS mode, typically 3 micro shutters are opened. The central shutter contains the target star, and the shutter above and below are for measuring the sky background. The actual number of measured spectra depends on the level of crowding in the field. In a highly dense region like 30 Doradus, there is an ample number of PMS stars to observe. The limiting factor comes from the number of contaminants in the field of view. A contaminant is a nearby star that falls within the 3 micro shutters being used to observe a target. If both the upper and lower shutter of the mini-slit contain contaminants, then it will not be possible to accurately measure the nebular background, in order to subtract it from the target spectrum. Similarly, if a contaminant falls within the same micro shutter as the target, then it is not possible to measure a reliable flux for the target. As such, careful consideration goes in to which targets are selected.\\
For this study, the cases where a contaminant occupies only 1 of either the upper or lower shutters were permitted, as at least one nebular background value could still be measured from the remaining `free' shutter. Naturally it is preferable if there are no contaminants at all. It is advised \cite{ferruit2022near} to carry out observations in the NIRSpec MOS mode using 3 nodding positions. The first nodding position places the target in the central shutter. The following nodding position then places the target in the upper shutter, and finally the third nodding position places the target in the bottom shutter. Each nod shifts the target by roughly 5 pixels on the detector in the dispersion direction. This helps to smooth out detector blemishes and residual flat fielding errors \cite{ferruit2022near}. This physical movement of the telescope means that additional contaminants may enter the field of view of the mini-slit, and so the MPT must be consulted regarding this before the final configuration is decided.\\ 
The real estate of the MSA is also partially limited by failed shutters. Failed closed shutters are stuck in the closed position, and cannot be used for observation. The impact of these shutters is minimal, due to the sheer number of other available shutters on the MSA. The failed open micro shutters provide more of an obstacle to planning an observation. Since they are stuck open, unwanted light enters through these shutters, becomes dispersed, and spreads over the detector in the dispersion direction. This means that failed open shutters prevent entire rows (where rows refers to shutters running along dispersion direction) of the MSA from being used. There is also a risk that the unwanted dispersed light may intersect with target spectra placed on nearby rows, due to the irregular (curved) spectra produced by NIRSpec. Figure \ref{fig:MSA_sky_strategy} shows a section of the configured MSA, highlighting the failed open and closed shutters, as well as the nodding pattern used in this simulation. A final consideration in deciding on the configuration of the MSA is the detector gap, which causes a break in the spectra of some stars, depending on their position on the detector, and which resolution is being used. In high resolution mode (R=2700) spectra will always encounter the detector gap, as the higher dispersion spreads the light over the entire detector. In low resolution mode (R=100) this is not necessarily true, as an entire spectrum can fit on one half of the detector. This study is focused on the emission lines of PMS stars. Depending on the MSA column that a target is placed on, it is possible for emission lines to fall within the detector gap. It is desirable to place stars on the MSA such that only the continuum falls within the detector gap, however, in some cases this is not possible. With all of these considerations, the number of target stars for 30 Doradus reduced from about 50 to about 40 for a given nodding position. The breakdown of targets that were discarded is summarised in table \ref{tab:good_primaries}.  

\begin{table}[h]
\centering
\scalebox{0.7} {
\begin{tabular}{|c|c|c|}
 \hline
 \multicolumn{3}{|c|}{Overview of targets per nodding position} \\
 \hline
 Nodding position & Total targets & Clean targets \\ [0.5ex] 
 \hline\hline
 1 & 51 & 43 \\ 
 \hline
 2 & 52 & 44 \\ 
 \hline
 3 & 50 & 42 \\ 
 \hline
\end{tabular} 
}
\caption{\label{tab:good_primaries}Table showing the reduction in the number of targets with at least one free background shutter.}
\end{table}

\subsection{Creating a count rate map.}
To simulate a NIRSpec MOS observation, both stellar spectra as well as the background spectrum were required. The stellar spectra used here were simple model spectra, based on the known stars of 30 Doradus that lie within a 3 arc minute window of the core. These models consisted of a theoretical spectrum, with emission lines added in the same way as for the nebular background spectrum, using the $H_{alpha}$ emission line luminosity measured for each star with Oosterbrock case B recombination. The stellar models were processed by the IPS, producing an electron rate map (ERM) of the NIRSpec detectors. This product is equivalent to the 2D spectra that would be produced during a real observation, with one key difference - the ERM is noiseless. Noise is introduced to the ERM by combining the photon noise, readout noise and dark current expected for a real observation. Adding this noise to the ERM transforms it into a count rate map (CRM). The same process was then applied to the background spectrum. The stellar CRM and background CRM were then added together, pixel by pixel. This combined CRM is the final data product, consisting of stellar spectra, with nebular background contamination, as well as noise.

\subsection{Reducing the data with NIPS}
The NIRSpec Instrument Pipeline Software (NIPS) is an automatic data reduction pipeline for NIRSpec data. NIPS performs spectrum extraction, background subtraction, flux calibration, and wavelength calibration. It is possible to specify which of the shutters should be used for measuring the nebular background, in the situation where a contaminant occupies one of the shutters. This step is accomplished by referring to the `planning file', a table that specifies where the targets, contaminants and fillers (additional stars that are added into the MSA if there is free space) lie on the MSA. In order to measure the nebular background from the appropriate shutters, a script was written that flags targets with contaminants in both background shutters, as well as targets in which only a single background shutter is unusable. In the former case, this target must unfortunately be skipped, as no background subtraction can take place. In the latter case, the single contaminated shutter was then removed from the list of available background shutters, leaving the free background shutter to be used for background subtraction. The background subtraction is performed by measuring the pixels that contain only nebular background, and calculating the median pixel value. Simply using the median background value assumes that the background does not vary significantly between shutters. This assumption is discussed in the following section.

\subsection{Spatial variation of the nebular background}
\begin{figure}[h!]
    \centering
    \begin{subfigure}{0.49\linewidth}
         \includegraphics[width=1\linewidth]{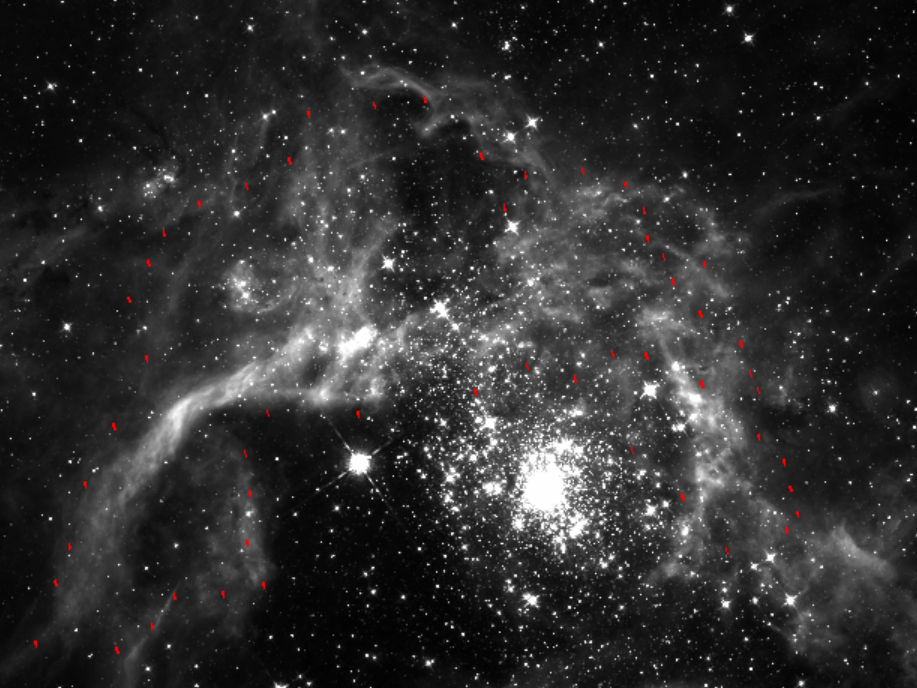}
    \end{subfigure}
 \hfill
    \begin{subfigure}{0.49\linewidth}
         \includegraphics[width=1\linewidth]{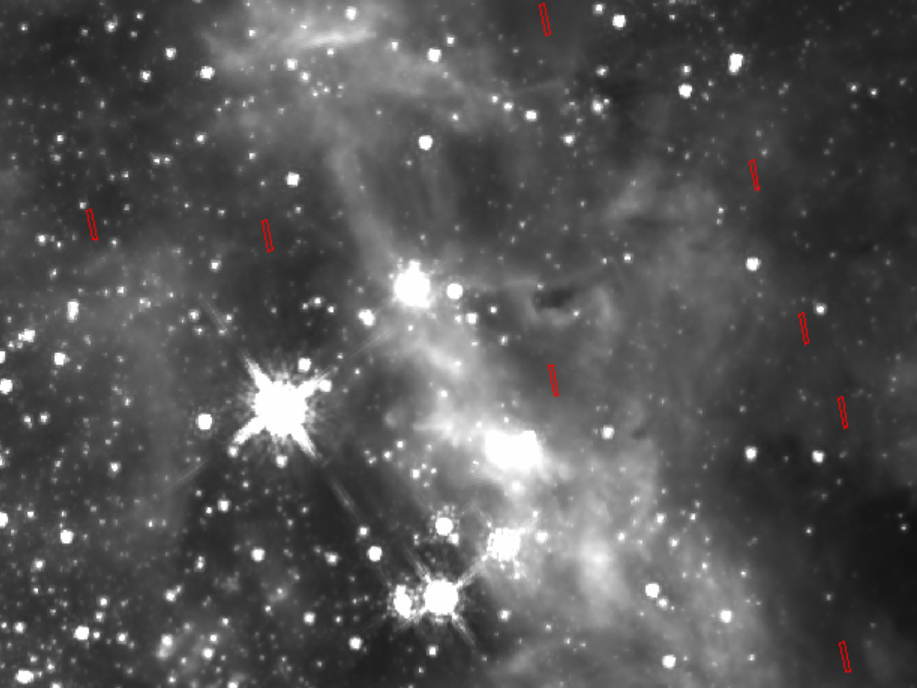}
     \end{subfigure}
  \caption{Left: $F160W$ image of 30 Doradus showing the manually placed mini-slits in and around the nebula. The position of the mini-slits can be limited to $336.15168149^{\circ}$ and corresponds to the roll angle that JWST will have when observing 30 Doradus. Right: Zoom in of several mini-slits from the same image. The background is clearly variable over larger scales, yet since the angular area of each mini-slit is small, the background remains approximately constant.}
  \label{fig:30_dor_regions}
\end{figure}

\begin{figure}[h]
    \centering
    \includegraphics[width=0.7\linewidth]{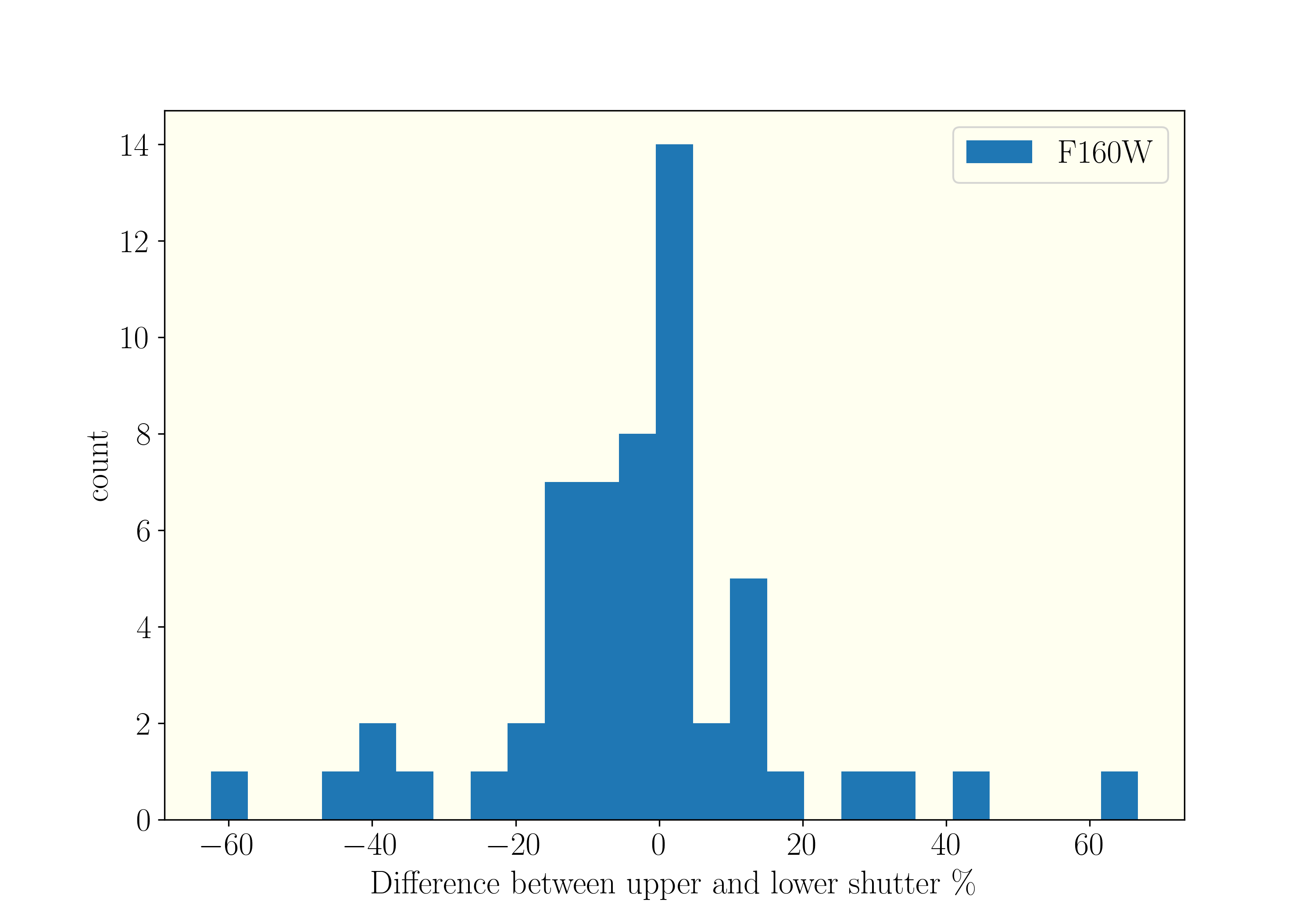}
    \caption{Histogram showing the typical degree of variation between upper and lower shutters in the $F160W$ filter.}
    \label{fig:bkg_variation_hist}
\end{figure}

An assumption was made at the beginning of the simulation, that the nebular background spectrum does not vary across the field of view. This includes changes in brightness, but also changes in the shape of the spectrum. This assumption greatly simplified the simulations. To investigate whether this assumption is valid for dense star formation regions like 30 Doradus, a test was performed on the HST images. This test consisted of measuring the nebular background, and investigating how the intensity varies across the angular scale of the mini-slit. The images used in this test were: $F658N$, $F110W$ and $F160W$. The two wide band filters were selected because they are within the same wavelength range as NIRSpec, and so provide immediate information about how the infrared continuum emission of the 30 Doradus nebula varies spatially. The narrow band $H_{\alpha}$ filter was selected as emission line strength varies not just as a function of temperature and metallicity, but also due to density changes, degree of ionisation, as well as radiation density of the medium \cite{byler2017nebular}. An aperture was created with the same physical dimensions as the mini-slit, i.e. 3 micro shutters, measuring $0.2"$ across and $1.38"$ in length. The aperture was placed at the roll angle of $336.15168149^{\circ}$, which is currently applied to the NIRSpec observations of 30 Doradus to be executed in 2023 \cite{de2017star}. As future observations of 30 Doradus will be made at different roll angles, a range of angles were tested. It was found that the roll angle of the telescope had no meaningful impact on the degree of spatial variation observed. The rectangular aperture was then placed manually at different locations in an image, ensuring that no stars were present in the mini-slit, so that only nebular background emission was being measured. The best image in the dataset to ensure that no stars are present in the mini-slit was $F160W$. This is because the optical depth of the nebula is significantly lower in the infrared compared to the optical, and hence stars that may have been obscured by nebular emission in the $F658N$ image, clearly emerge in the $F110W$ image, and more so in the $F160W$ image. Placing the shutters manually in the image provided a reliable way to avoid contaminant starlight, but also limited the number of locations that could be sampled before becoming prohibitively time consuming. The mini-slit was divided into 3 equal sections, each section representing a single micro shutter. The nebular emission in the upper and lower shutters was measured and the two values were compared. This was repeated 50 times, sampling a different region of the image each time. This was done for the 3 photometric images. The aperture locations were the same for all 3 images. On the right of figure \ref{fig:30_dor_regions} is the $F160W$ image of 30 Doradus with the 50 manually sampled regions. On the left is a close up more clearly showing the individual mini-slits. Regions near the core of the star forming region as well as outwards towards the edge of the nebula were sampled, including areas that visually appeared to vary in brightness over small spatial scales. Figure \ref{fig:bkg_variation_hist} shows the distribution of differences in brightness between the upper and lower shutters. For the $F160W$ image, the background typically varied by $3\%$ between the shutters, with a $1\sigma$ spread of $20\%$. For $F110W$, the background typically varies by $1 \%$ between the upper and lower shutters with a $1\sigma$ spread of $20\%$. For $F658N$, the background typically varies by $2 \%$ between the upper and lower shutters with a $1\sigma$ spread of $22\%$. The variation being highest for $F658N$ was expected based on the additional physical properties that influence line emission. However, the small typical difference between shutters and narrow spread indicates that a non-varying background is a reasonable assumption for these simulations, both for continuum and for line emission.

\section{Results of the MOS simulation}
\subsection{Assessing the performance of NIPS background subtraction on emission lines.}
\begin{figure}[h]
    \centering
    \includegraphics[width=1\linewidth]{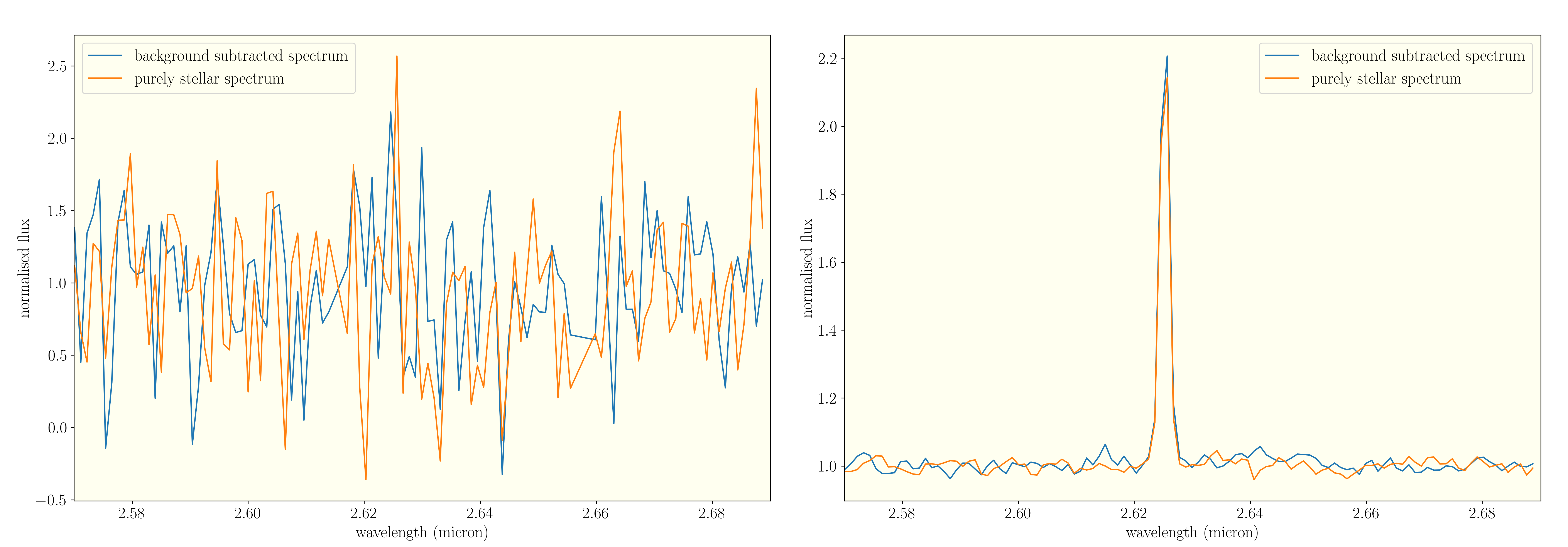}
    \caption{Left: Example of a noisy emission line with signal-to-noise ratio of $2$, and a likely overestimated EW of $103 \AA$. Right: Example of a prominent emission line with a signal-to-noise ratio of $16$, and an EW of $101 \AA$. The blue line is that of the background subtracted stellar line. The orange line is that of the purely stellar line, before any background contamination had been added.}
    \label{fig:noisy+clean_emission_line}
\end{figure}

In order to assess the performance of the background subtraction, the EW of the 3 prominent stellar emission lines: $Pa_{\alpha}$, $Br_{\beta}$ and $Br_{\gamma}$ were measured. The background emission will add energy to these lines, increasing the EW. This means that the emission lines consist of both a stellar and a nebular component. The aim of the background subtraction is to remove the nebular component, leaving the stellar part intact. Stellar and nebular emission lines differ greatly in terms of how they are affected by broadening mechanisms. Nebular lines are extremely narrow, due to lack of thermal, rotational, or collisional broadening. PMS stars on the other hand have significantly broader emission lines, due to the high temperatures that matter is heated to as it  accretes onto the star \cite{hartmann2016accretion}. The medium resolution dispersion grating used in this simulation means that neither the narrow nebular lines, nor the broader stellar lines are resolved. Both simply take on the profile of the line spread function. This actually makes removal of the background simpler, as the shape of the lines is the same, with only the brightness/intensity of the lines differing. The EW of the lines was measured here by fitting a linear function to the continuum around each emission line, in order to estimate the level of the continuum. The area of the emission lines was then measured using the trapezoidal method. The area of each emission line was then divided by their respective continuum level, resulting in the EW of each line. Using the same approach, the EW of the emission lines in the adopted stellar models were measured, and compared to the EW of the background subtracted simulated spectra. Because 3 nodding positions were used, the total number of stellar spectra roughly tripled from about 40 to 120, giving about 360 individual emission lines. The final number of emission lines used in the analysis was less than this, due to the detector gap, as well as the exclusion of some lines that featured extremely low signal-to-noise. A threshold of $\frac{S}{N}=4$ was used here. This threshold was decided based on when an emission line was clearly distinguishable above the continuum. Figure \ref{fig:noisy+clean_emission_line} shows the $Br_{\beta}$ emission line for two different targets in the same nodding position. The blue line has been background subtracted, and the orange line is the clean, nebular free stellar line. On the left of figure \ref{fig:noisy+clean_emission_line} is an example of a noisy source, below the threshold of $\frac{S}{N}=4$, and as such this particular line was excluded from analysis. $Br_{\beta}$ was typically the weakest emission line within the wavelength range used here, and so even if it was excluded, it is possible that $Pa_{\alpha}$, or $Br_{\gamma}$ could be measured for the same source. On the right of figure \ref{fig:noisy+clean_emission_line} is an example of a prominent, high signal-to-noise source, in which the EW could be accurately measured.

\section{Results}
\begin{figure}[h]
    \centering
    \includegraphics[width=0.7\linewidth]{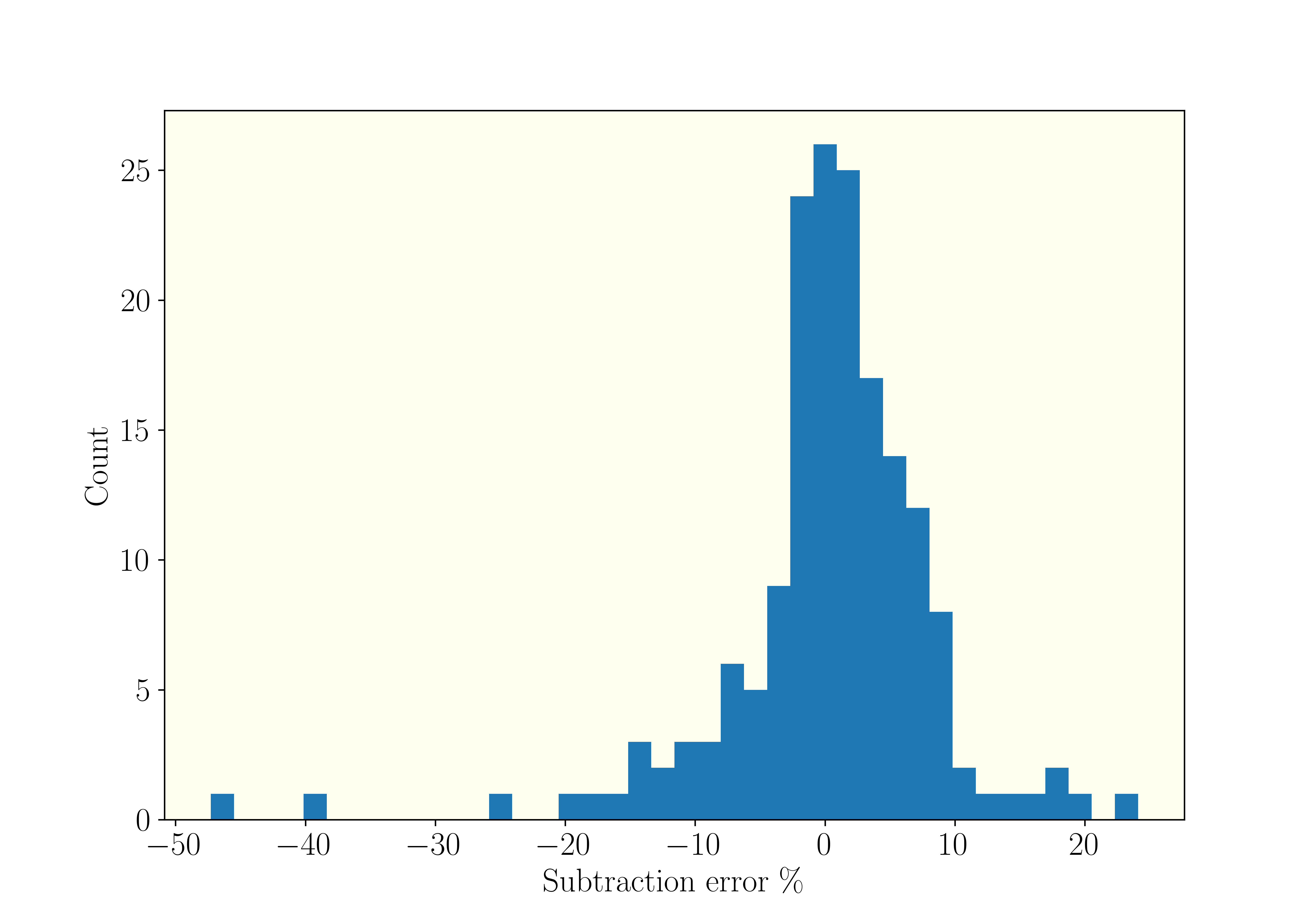}
    \caption{The distribution of nebular emission subtraction errors from the emission lines $Pa_{\alpha}$, $Br_{\beta}$ and $Br_{\gamma}$. }
    \label{fig:bg_sub_errors}
\end{figure}

Of the roughly 40 targets that could be placed within the simulated MSA configuration, a small number had to be discarded due to cases in which one of the three emission lines fell within the detector gap, meaning that the measurement of the EW of that particular line was not possible.  A total of 223 stellar emission lines were used to assess the performance of NIPS' background subtraction. Figure \ref{fig:bg_sub_errors} shows the distribution of errors of the background subtraction. It was found that the typical difference in EW of a background subtracted line compared to the original model was $0.8\%$, with a $1\sigma$ spread of $13\%$.

\subsection{Discussion}
The typical subtraction error of $<13\%$ indicates that NIPS' background subtraction is able to reliably return an accurate EW for strong stellar emission lines. Because of the approximation used in this simulation of a spatially constant background spectrum, the actual uncertainty is likely slightly larger. However, even errors on the order of $50 \%$ would not prevent a reliable determination of the mass accretion rate, as there are other uncertainties that dominate the total uncertainty of measuring the mass accretion rate of PMS stars. This includes estimation of stellar parameters like mass and temperature, as well as extinction and bolometric correction, in order to convert from a total accretion luminosity to a mass accretion rate \cite{herczeg2008uv}. Emission lines themselves are also not perfect probes of accretion rate. Other stellar processes can produce emission lines such as stellar winds, which are unrelated to mass accretion from the disk \cite{kraus2008detection}. It is noted again that for this simulation, a region of particularly high nebular brightness was chosen in order to assess a `worst case scenario'. The actual observations will likely suffer less from nebular contamination over all. It is also possible to refine the nebular subtraction approach itself. Calculating a median value between the upper and lower shutters is certainly a first order attempt. Here it has been shown to work reasonably well, as the background does not change considerably over the angular scale of the mini-slit for 30 Doradus. In star forming regions where this may no longer be true, more refined estimations of the background level can be taken by fitting polynomial functions to the background level, which would account for rapid changes in brightness. A future challenge that has not been addressed yet is the removal of nebular contamination when using the high resolution mode (R=2700) of NIRSpec. In high resolution mode, the stellar emission lines will be resolved, while the nebular lines will not. This will mean that the profile of the nebular line and stellar line will no longer be the same, and performing a simple subtraction pixel by pixel will not adequately remove the line. In these cases, a larger residual is expected in the subtracted stellar emission lines. Thankfully this will mostly be in the core of the line, while the wings, which contain most of the kinematic information of the gas, are mostly unaffected. This means that physical parameters such in the infall velocity of material from the disk to the star can be determined in high resolution mode, while medium or even low resolution mode can be used to determine mass accretion rate based solely on the EW of the lines.

\subsection{Summary}
The background subtraction for NIRSpec's MOS mode has been assessed by simulating a MOS observation in the crowded and bright star forming region of 30 Doradus, using the medium resolution mode between $1.6 - 3 \mu m$. A model spectrum of the nebular background was created based on HST photometry of 30 Doradus at optical and near-infrared wavelengths. The EW of stellar emission lines was measured on the simulated extracted spectra, before and after the addition of nebular contamination and a typical subtraction error of $0.8\%$ was found, with a $1 \sigma$ spread of $13\%$. The choice of a spatially non-variable background is justified based on the measurement of typical brightness changes at $\lambda = 1.6 \mu m$ of $0.05\%$ across the angular scale of the mini-slit in the densest region of 30 Doradus. The automated data reduction pipeline - NIPS, performs remarkably well in dense star formation regions, despite having been designed with galaxy observations in mind, where only the weak continuum of the zodiacal light needs to be removed. It is likely only the faintest sources, in the brightest nebulae that will require special attention in order to accurately remove the nebular contamination.

\bibliography{references} 

\begin{thebibliography}{10}

\bibitem{byler2017nebular}
Byler, N., Dalcanton, J.~J., Conroy, C., and Johnson, B.~D., ``Nebular
  continuum and line emission in stellar population synthesis models,'' {\em
  The Astrophysical Journal}~{\bf 840}(1),  44 (2017).

\bibitem{walborn1990contemporary}
Walborn, N.~R. and Fitzpatrick, E.~L., ``Contemporary optical spectral
  classification of the ob stars: a digital atlas,'' {\em Publications of the
  Astronomical Society of the Pacific}~{\bf 102}(650),  379 (1990).

\bibitem{dahm2008spectroscopic}
Dahm, S., ``A spectroscopic examination of accretion diagnostics for near solar
  mass stars in ic 348,'' {\em The Astronomical Journal}~{\bf 136}(2),  521
  (2008).

\bibitem{sabbi2013hubble}
Sabbi, E., Anderson, J., Lennon, D., Van Der~Marel, R., Aloisi, A., Boyer,
  M.~L., Cignoni, M., De~Marchi, G., De~Mink, S., Evans, C., et~al., ``Hubble
  tarantula treasury project: Unraveling tarantula's web. i. observational
  overview and first results,'' {\em The Astronomical Journal}~{\bf 146}(3),
  53 (2013).

\bibitem{sabbi2016hubble}
Sabbi, E., Lennon, D., Anderson, J., Cignoni, M., Van Der~Marel, R., Zaritsky,
  D., De~Marchi, G., Panagia, N., Gouliermis, D., Grebel, E., et~al., ``Hubble
  tarantula treasury project. iii. photometric catalog and resulting
  constraints on the progression of star formation in the 30 doradus region,''
  {\em The Astrophysical Journal Supplement Series}~{\bf 222}(1),  11 (2016).

\bibitem{ferruit2022near}
Ferruit, P., Jakobsen, P., Giardino, G., Rawle, T., de~Oliveira, C.~A.,
  Arribas, S., Beck, T., Birkmann, S., B{\"o}ker, T., Bunker, A., et~al., ``The
  near-infrared spectrograph (nirspec) on the james webb space telescope-ii.
  multi-object spectroscopy (mos),'' {\em Astronomy \& Astrophysics}~{\bf 661},
   A81 (2022).

\bibitem{guido_30dor}
De~Marchi, G. and Panagia, N., ``The extinction law inside the 30 doradus
  nebula,'' {\em Monthly Notices of the Royal Astronomical Society}~{\bf
  445}(1),  93--106 (2014).

\bibitem{melnick198530}
Melnick, J., ``The 30 doradus nebula. i-spectral classification of 69 stars in
  the central cluster,'' {\em Astronomy and Astrophysics}~{\bf 153},  235--244
  (1985).

\bibitem{osterbrock1981seyfert}
Osterbrock, D.~E., ``Seyfert galaxies with weak broad h alpha emission lines,''
  {\em The Astrophysical Journal}~{\bf 249},  462--470 (1981).

\bibitem{crowther2018dissecting}
Crowther, P.~A., Castro, N., Evans, C., Vink, J., Melnick, J., and Selman, F.,
  ``Dissecting the core of the tarantula nebula with muse,'' {\em arXiv
  preprint arXiv:1801.00855}  (2018).

\bibitem{de2017star}
De~Marchi, G., Alves~de Oliveira, C., Beck, T., Biazzo, K., Ferruit, P.,
  Giardino, G., Maiolino, R., Muzerolle, J., Panagia, N., and Sabbi, E., ``Star
  formation in the local group-ngc 2070 (30 dor),'' {\em JWST Proposal. Cycle
  1} ,  1226 (2017).

\bibitem{hartmann2016accretion}
Hartmann, L., Herczeg, G., and Calvet, N., ``Accretion onto pre-main-sequence
  stars,'' {\em Annual Review of Astronomy and Astrophysics}~{\bf 54},
  135--180 (2016).

\bibitem{herczeg2008uv}
Herczeg, G.~J. and Hillenbrand, L.~A., ``Uv excess measures of accretion onto
  young very low mass stars and brown dwarfs,'' {\em The Astrophysical
  Journal}~{\bf 681}(1),  594 (2008).

\bibitem{kraus2008detection}
Kraus, S., Preibisch, T., and Ohnaka, K., ``Detection of an inner gaseous
  component in a herbig be star accretion disk: near-and mid-infrared
  spectrointerferometry and radiative transfer modeling of mwc 147,'' {\em The
  Astrophysical Journal}~{\bf 676}(1),  490 (2008).

\end{thebibliography}
\bibliographystyle{spiebib} 

\end{document}